\documentclass[12pt,preprint]{aastex}
\received{31 December 2002}
\accepted{6 March 2003}
\slugcomment{ApJ, 590, 20 June 2003 (scheduled)}

\shorttitle{Abundance Analysis of IRAS 06530$-$0213}
\shortauthors{Hrivnak \& Reddy}

\begin{document}

\title{AN ABUNDANCE ANALYSIS OF THE NEW CARBON-RICH
PROTO-PLANETARY NEBULA IRAS~06530$-$0213}

\author{Bruce J. Hrivnak\altaffilmark{1}}
\affil{Department of Physics and Astronomy, Valparaiso University,
Valparaiso, IN 46383; bruce.hrivnak@valpo.edu}
\and
\author{Bacham E. Reddy}
\affil{Department Astronomy, University of Texas, Austin, TX 78712; 
ereddy@astro.as.utexas.edu}

\altaffiltext{1}{Visiting astronomer, Kitt Peak National Observatory,
National Optical Astronomy Observatories, which is operated by the
Association of Universities for Research in Astronomy, Inc., under
contract with the National Science Foundation.}

% The abstract environment prints out the receipt and acceptance dates
% if they are relevant for the journal style.  For the aasms style, they
% will print out as horizontal rules for the editorial staff to type
% on, so long as the author does not include \received and \accepted
% commands.  This should not be done, since \received and \accepted dates
% are not known to the author.

\begin{abstract}

In this paper, we present a study of the proto-planetary nebula (PPN)
IRAS~06530$-$0213 based on low- and high-resolution spectra.
The low-resolution spectrum shows that star is an F5 supergiant with
molecular C$_{2}$ and C$_{3}$ and enhanced s-process lines.
From the high-resolution spectra,
the following atmospheric parameters were determined:
$T_{\rm eff}$~=~6900~K, log~$g$~=~1.0 and $\xi_{\rm t}$~=~4.5.
Abundance analysis shows that IRAS~06530$-$0213 is metal-poor
([Fe/H]=$-$0.9) and overabundant in carbon ([C/Fe]=1.3),
nitrogen ([N/Fe]=1.0), and s-process elements ([s-process/Fe]=1.9),
indicating AGB nucleosynthesis and deep convective mixing.
From the analysis of circumstellar C$_{2}$ and CN molecular bands in
the spectrum of IRAS~06530$-$0213, an envelope expansion velocity
of V$_{\rm exp}$=~14$\pm1$~km s$^{-1}$
was determined, a typical value for post-AGB stars.
Also typical of PPNs is the double-peaked spectral energy distribution.
The properties of both the photosphere and circumstellar envelope
suggest that IRAS~06530$-$0213 is unambiguously a low-mass, carbon-rich PPN.
For comparison purposes, new, high-resolution spectra of the well-known
PPN HD~56126 (IRAS~07134+1005) was also analyzed and compared
with previous results.

\end{abstract}


\section{INTRODUCTION}

Proto-planetary nebulae (PPNs) represent an important transitional phase
in the evolution of low and intermediate-mass stars, between the asymptotic
giant branch (AGB) and the planetary nebula (PN) phases.
During this phase, the circumstellar envelope of gas and dust is detached and
expanding away from the star.  The star, meanwhile, is increasing in surface
temperature as it moves approximately horizontally (constant luminosity)
across the H-R diagram.
This evolutionary phase is expected to last several thousand years,
depending upon the mass of the star \citep{blocker95}.
When the temperature of the star becomes high enough to significantly
photoionize the gas (T$_{\star}$ $>$ 30,000~K), an emission-line spectrum is
seen and the star has passed to the PN phase.
However, when still in the PPN phase, the nebula is small and seen only
in reflected light.

The study of PPNs has grown over the past two decades,
following the identification of PPN candidates based upon their infrared
excesses in the {\it IRAS} database.  (The infrared emission arises from the
absorption and re-emission of radiation from their circumstellar dust.)
The PPN/post-AGB nature of these stars is supported by several
lines of evidence.  These include the following.  (1) The spectral
energy distribution (SED), based upon the visible, near-infrared, and
mid-infrared flux measurements, reveals a double-peaked
SED with approximately equal amounts of flux received from the
reddened photosphere and the warm dust (T$_{\rm d}$$\sim$150$-$250~K).
(2) Millimeter spectral-line observations (CO or OH) reveal an
expanding gas envelope with V$_{\rm exp}$$\sim$10~km~s$^{-1}$.
(3) Low-resolution spectra show the expected spectral and
luminosity class (B$-$G supergiants).
(4) High-resolution spectra show the expected evidence of
nucleosynthesis during the AGB phase of evolution.
A description of the discovery and basic properties of PPNs is given
by \citet{kwok93}, with recent updates from higher-resolution studies
discussed by \citet{kwok00} and \citet{hrivnak03}.

The discovery of molecular carbon, C$_{2}$ and C$_{3}$, in the blue region
of the visible spectrum \citep{hrivnak95} and C$_{2}$ and CN in the red region
\citep{bakker96} of several PPNs indicated that they were carbon-rich.
Interestingly, these same carbon-rich PPNs also have been found to
display an unidentified feature in their mid-infrared spectra, the
so-called ``21 $\mu$m feature'' \citep{kwok89, kwok99}.

Quantitative abundance studies of the these PPNs with molecular carbon
have determined that they are somewhat metal-poor, with an
overabundance of CNO and a large overabundance of the
s-process elements such as Y, Zr, La, Ce, and Nd \citep{vanwinckel00, reddy02}.
These studies provide strong evidence that these objects are post-AGB stars
that have undergone nucleosynthesis and convective mixing during the AGB
phase.

In this paper, we report on our study of the new carbon-rich PPN,
IRAS~06530$-$0213.
It's likely identification as a PPN was
reported by \citet{hu93}, who classified it as FO~I with V=14.
Our photometry and low-resolution
spectroscopy confirm the basic results of \citet{hu93} as to its PPN
nature, but they also reveal for the first time the presence of C$_2$ and C$_3$.
New high-resolution spectra were obtained to determine detailed
abundances of the elements and confirm the post-AGB nature
of the star.
On the basis of its SED, spectral classification, elemental abundances,
and circumstellar envelope, IRAS~06530$-$0213 is shown unambiguously
to be a low-mass PPN.
Also included is the analysis of new high-resolution spectra of the
relatively well-known, bright (V=~8.2) PPN HD~56126
(IRAS~07134+1005; F5~I).
Although it has been the subject of previous abundance analyses,
we have included it for the purpose of comparison with the results of
IRAS~06530$-$0213.
Preliminary results of this study have been presented by \citet{reddy99a}.

\section{OBSERVATIONS}

\subsection{Visible and Infrared Photometry}

We had initially selected IRAS 06530$-$0213 as a PPN candidate based
upon its {\it IRAS} colors, which peaked in the 25 $\mu$m bandpass.
Correlation with the STScI Guide Star Catalog revealed a 14th mag
star close to the {\it IRAS} position.  The association with the
{\it IRAS} source was confirmed by a ground-based 10 $\mu$m
observation carried out on the United Kingdom Infrared Telescope
on Mauna Kea under their Service Observing program.
A measurement of N $=$ 2.32 $\pm$ 0.04 was  made on 1993 May 6.

Standardized photometry of IRAS 06530$-$0213 was carried out at
Kitt Peak National Observatory (KPNO).
Visible light observations were made using a CCD on the
0.9~m telescope on 1995 September 11 and
near-infrared observations were made using the
Simultaneous Quad-color Infrared Imaging Device (SQIID)
on the 1.3~m telescope on 1993 November 9 and 10.
SQIID used a dichroic crystal to separate the incoming beam so that
it could be observed in four infrared bandpasses simultaneously.
The resulting photometric values are listed in Table \ref{ptm}.
They are in good agreement with the previously published photometric
values of \citet{hu93}, \citet{reddy96}, and \citet{gl97}.
The object appears very red, with (B$-$V) = 2.3.

\placetable{ptm}

\subsection{Low-Resolution Spectroscopy}

A low-resolution spectrum of IRAS~06530-0213 was obtained on
1992 October 09 with the Gold Camera Cassegrain Spectrograph on the
2.1~m telescope at KPNO.
The spectrum had a wavelength range of 3850$-$7600 \AA\  and
had a resolution of 8.0 \AA.
The spectrum was reduced with IRAF\footnote{IRAF is distributed by the 
National Optical Astronomical
Observatory, which is operated by the Association of Universities for
Research in Astronomy, Inc., under contract to the National Science
Foundation.} using standard reduction procedures -
bias subtraction, flat-field correction, sky subtraction, one-dimensional
extraction, and wavelength calibration using an arc lamp (He-Ne-Ar).
Several spectral standards were also observed.

\subsection{High-Resolution Spectroscopy}

High-resolution spectra of IRAS~06530$-$0213 were obtained with the 
2.7~m telescope at McDonald Observatory on 1997 October 17 (R$\approx$55,000) 
and 2001 December 10 (R$\approx$45,000).
The telescope was equipped with a cross-dispersed Coud\'{e}
echelle spectrograph \citep{tull95} and a CCD (2048$\times$2048 pixels).
%Since the projected size of the echelle spectrogram is larger than the size
%of the CCD, spectral gaps appear  between the orders.
Three individual spectra of 45 minutes each were acquired on each night,
and the spectra on each night were combined to remove cosmic ray hits and 
to improve the signal-to-noise ratio (S/N). 
For the purpose of wavelength calibration, Th-Ar comparison spectra
were acquired before and after each of the sets of program star exposures.

IRAS 06530$-$0213 is faint (V=14.1) and highly reddened.  
Because of this, only a moderate S/N of 30 at 6560~\AA\ was achieved on the
earlier date and a higher S/N of 80 at 6560~\AA\ at slightly lower resolution 
was achieved on the later date.
We could confidently only use the spectra in the region between
5200$-$9200~\AA\ but not at shorter wavelengths where the signal
decreases rapidly.
A hot, rapidly rotating bright star was also observed for use in
distinguishing telluric lines from the stellar features.

High-resolution spectra of HD~56126 were obtained on 1996 December 23 
(t$_{\rm exp}$=30 min, R$\approx$55,000) and on 2002 November 14 
(2 spectra of t$_{\rm exp}$=30 min, R$\approx$55,000)
with the same telescope and equipment.
The observations cover the wavelength range 4000$-$9800~\AA\, with gaps
between echelle orders, and they result in spectra with S/N ranging from
100 to 400.

These high-resolution spectra were reduced using IRAF.
Standard reduction procedures were employed, which consisted of bias
subtraction,
flat-field image correction, extraction of one-dimensional spectra, wavelength
calibration, and continuum fitting.

\section{LOW-RESOLUTION SPECTRUM: CLASSIFICATION}

\begin{figure}
\figurenum{1}
\caption{Low-resolution spectrum of IRAS~06530$-$0213, along with
two spectral classification standard stars. 
The spectra are not flux-calibrated.  
Note the presence of molecular features and also weak Balmer 
lines in the spectrum of IRAS~06530$-$0213.}
\label{low-res}
\plotone{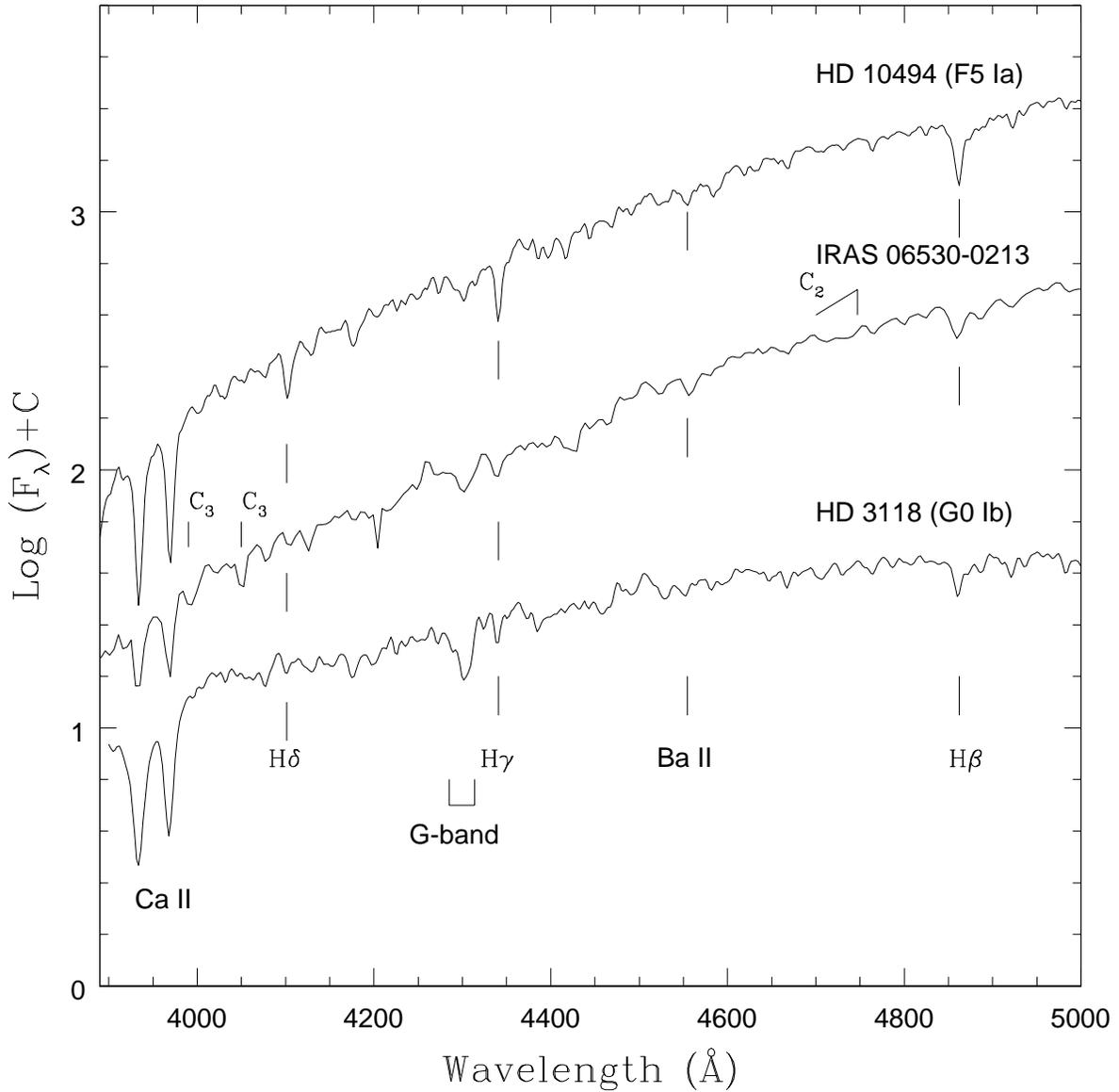}
\end{figure}

The extracted blue part (3800$-$5000~\AA) of the low-resolution spectrum  
of IRAS~06530$-$0213 is displayed in Figure~\ref{low-res}, 
along with the spectra of two supergiants shown for comparison.
The y-axis is plotted on a logarithmic scale to display the weaker
features in the blue more clearly.
A few prominent spectral features are identified in the figure.
Of particular importance is the identification of strong molecular
C$_{2}$ (4737~\AA) and C$_{3}$ (3991~\AA, 4050~\AA) features.
Also seen is a strong Ba\,{\sc ii} feature at 4554~\AA.
These enhanced molecular and s-process features are commonly seen in
carbon-rich PPNs (Hrivnak 1995).

The weak CH-band at 4300~\AA\ (G-band) suggests that the spectral type is
earlier than G0, and the ratio of the strengths of the
Ca\,{\sc ii} H line and the Ca\,{\sc ii} K + H$\epsilon$ lines ($\sim$1.0)
indicates that it is not earlier than F0.
From comparison with spectral standards, we assign the spectral
classification F5~I.
As seen in Figure~\ref{low-res},
the Balmer profiles (H$\beta$, H$\gamma$, and H$\delta$)
and metallic lines, such as Fe at 4045 \AA\ and Ca\,{\sc ii} H and K,
are weaker than expected in a star of spectral class between F5 and G2.
This may be due to some emission infilling of the Balmer profiles and to
a low abundance of metals.
\citet{hu93} and \citet{reddy96} had classified the object F0~I, but that
appears to be somewhat too early a spectral type for our spectrum.
From the analysis of the low-resolution spectrum, we qualitatively describe
IRAS~06530$-$0213 as a metal-poor, F5 supergiant with enhanced carbon
and s-process elements. These characteristics suggest that it is indeed
a post-AGB star, consistent with the assignment as a PPN.

\placefigure{low-res}

\section{ANALYSIS OF HIGH-RESOLUTION SPECTRUM}

\subsection{Description of Spectrum}
\begin{figure}
\figurenum{2}
\caption{Sample spectra of IRAS~06530$-$0213 and HD~56126
showing the strength of a few carbon and $s$-process lines.}
\label{high-resS}
\plotone{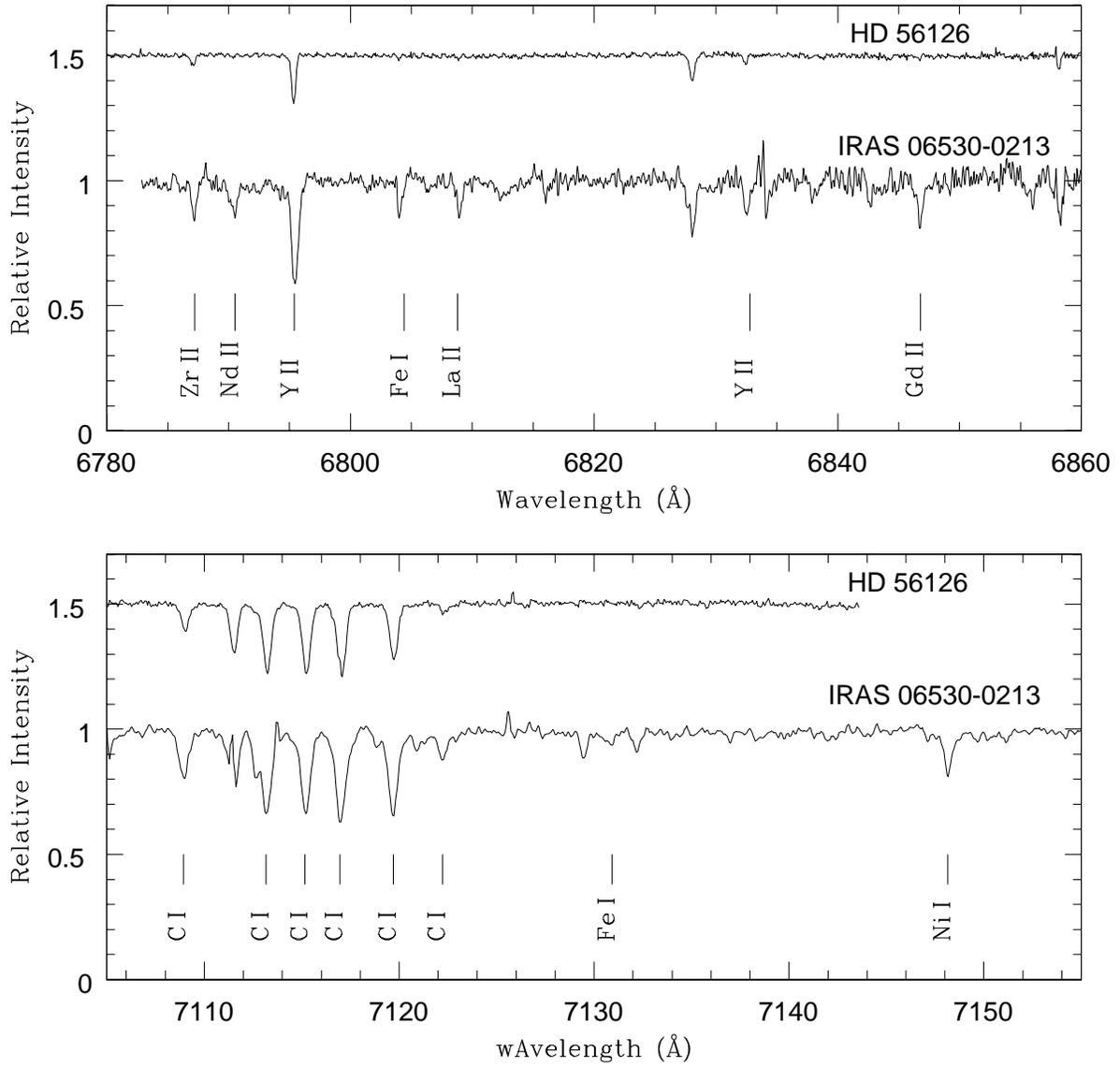}
\end{figure}

The high-resolution spectrum of IRAS~06530$-$0213 resembles
that of HD~56126. They both possess sharp spectral features and weak
metallic lines, suggesting that IRAS~06530$-$0213 is also
an evolved, metal-poor star.
In Figure \ref{high-resS} are displayed continuum-fitted,
normalized regions of the spectra of both the stars,
showing the presence of strong carbon lines
as well as strong lines of s-process elements.
The spectra of both stars exhibit strong and complex profiles of
H$\alpha$, Na~D, and K\,{\sc i}, with H$\alpha$ displaying emission
components in both stars.
Absorption bands of C$_{2}$ Phillips and CN Red system are also present
in both; these are displayed and discussed in detail later in this study.
The C$_{2}$ Swan band at 5165~\AA\ is also seen in HD~56126.

\placefigure{high-resS}

\subsection{Spectral-Line Analysis}

Atomic line identification was done using the solar spectrum
\citep{moore66}.
In the spectra, 99 lines were identified for IRAS~06530$-$0213 and
207 lines were identified for HD~56126 for which the equivalent widths
could be measured confidently and which also have reliable atomic parameters.
The reason for the fewer lines measured for IRAS~06530$-$0213 is the
significantly lower S/N of the spectrum, especially at the shorter wavelengths,
due to its relative faintness.
A large number of neutral carbon lines
(19 for IRAS~06530$-$0213 and 33 for HD 56126) were measured.
Very few metallic lines are seen in the spectra, which is presumably due
to both the mid-F spectral types and also a real deficiency in metals
(see Sect. \ref{abundance}).
We tried to measure as many lines as possible that were common to both
spectra.  However, this was restricted by the difference in their spectral
quality and also by the gaps in the spectra.

Radial velocities were measured using many lines of C\,{\sc i}, O\,{\sc i},
N\,{\sc i}, Fe\,{\sc i}, and Fe\,{\sc ii}. 
This resulted in heliocentric velocities for IRAS~06530$-$0213 of
V$_{r}$~=~51.0$\pm$1.0 km s$^{-1}$ (1997 October 17) 
and V$_{r}$~=~50.4$\pm$1.0 km s$^{-1}$  (2001 December 10).
The radial velocities measured for IRAS~06530$-$0213 from photospheric atomic
lines are in good agreement with the value measured from millimeter CO
measurements, V$_{r}$~=~50 km s$^{-1}$ \citep[V$_{lsr}$=33 km s$^{-1}$;][]{hu94}.

From the two spectra of HD~56126, we measured
V$_{r}$~=~83.3$\pm$0.5 km s$^{-1}$ (1996 December 23) and
$V_{\rm r}$ = 86.2$\pm$0.5 km s$^{-1}$ (2002 November 14).  
HD\,56126 is known to undergo photospheric pulsations with radial 
velocities of 81.7 to 91.8 km s$^{-1}$ \citep{lebre96}.
Our measured values of $V_{\rm r}$ fall in this range.

Equivalent widths (W$_{\lambda}$) were measured by fitting Gaussian
profiles to the observed lines.
Measurements were made of 15 elements
(C, N, O, Mg, Si, Ca, Ti, Fe, Y, Zr, Ba, La, Ce, Pr, and Nd) for IRAS~06530$-$0213 and of
19 elements (C, N, O, Mg, Si, S, Ca, Sc, Ti, Cr, Fe, Y, Zr, Ba, La, Ce, Nd, Sm, and Eu)
for HD~56126.

Oscillator strengths ($gf$-values) were taken from various sources.
For Fe\,{\sc i} and Fe\,{\sc ii} transitions, the $gf$-values critically reviewed
by \citet{lambert96} were adopted and for C,N, and O, the $gf$-values were
taken from compilation by \citet{wiese96}.
For rest of the lines used in this study, 
$gf$-values were taken from compilations of either the 
NIST\footnote{See the web site at http://physics.nist.gov/cgi-bin/AtData.}
or the VALD\footnote{See the web site at http://www.astro.univie.ac.at.}
database.

\subsection{Atmospheric Parameters}

The model atmospheric parameters of effective temperature ($T_{\rm eff}$)
and surface gravity (log $g$) can in principle be estimated from photometry
and from spectral classification.
However, in practice these two methods have often proven ineffective in
the case of evolved post-AGB stars.
In many cases these stars are highly reddened, and it is not
possible to determine the precise contributions of the interstellar and
circumstellar components to derive the intrinsic stellar colors.
Spectral classification also may lead to erroneous temperatures due to  
the metal-weak nature of the stars and the difficulty in determining the 
continuum in the blue part of the spectrum in the presence of strong
molecular features.
Inspection of Figure~\ref{low-res} shows that the Balmer profiles of 
IRAS~06530$-$0213 are weak for its assigned F5~I spectral class.  
It is not clear whether 
this weakness is due to emission filling (H$\alpha$ shows emission) or
to a real deficiency in hydrogen content.  If the stars are H-poor,
then the standard model analysis may not be applicable.
In fact, there are stars (e.g., R CrB type) which are thought to be in a 
post-AGB evolutionary phase but with severely H-depleted (or He-rich)
atmospheres  \citep[see][]{pandey01}.  These stars presumably evolved 
from main sequence stars with normal H content.  
Thus it is possible that IRAS~06530$-$0213 is in fact mildly H poor, but 
in the absence of He lines (expected to be seen only at higher temperatures) 
we cannot determine this.
In the case of these two stars, their spectral classes were only used to 
estimate initial values for the temperature and gravity.

\begin{figure}
\figurenum{3}
\caption{Determination of $T_{\rm eff}$ and log $g$ for
IRAS 06530$-$0213 using $T_{\rm eff}$-log $g$ diagrams.
The equal abundance curves of Fe and Ca intersect at
($T_{eff}$, log $g$)~=~(6700~K, 1.3) and (6850~K, 0.2) for
IRAS~06530$-$0213 and HD~56126, respectively.}
\label{T-g}
\plotone{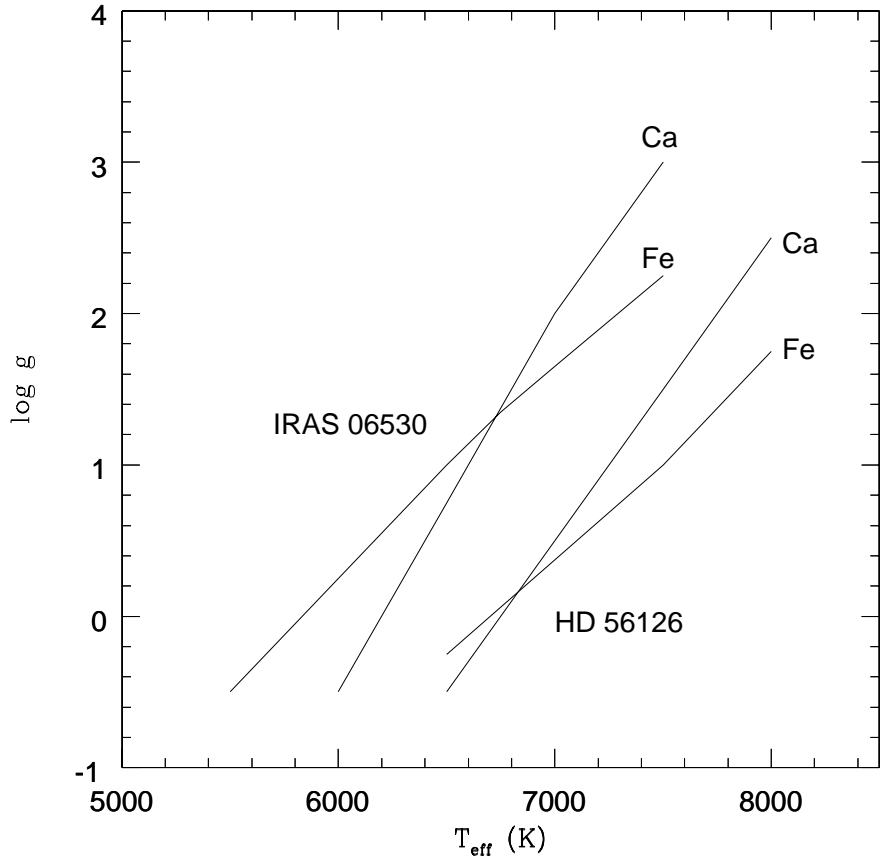}
\end{figure}

The atmospheric parameters $T_{\rm eff}$ and log~$g$ were first derived
using a $T_{\rm eff}$-log $g$ analysis.
In this method, we computed abundances for neutral and singly ionized
species of Fe and Ca for different values of $T_{\rm eff}$ and log $g$.
A value for the microturbulent velocity of $\xi_{t}$~=~5 km s$^{-1}$ was
assumed, which is a typical value for evolved supergiants.
In Figure~\ref{T-g} are plotted curves of equal abundances of
Fe\,{\sc i} and Fe\,{\sc ii} and of Ca\,{\sc i} and Ca\,{\sc ii} for each star.
The location of the intersection of the two curves yields the values of
$T_{\rm eff}$ and log $g$ and also the abundances of Fe and Ca for each star.
From this analysis we find for IRAS~06530$-$0213, $T_{\rm eff}$~=~6700~K
and log $g$~=~1.3,
and for HD~56126, $T_{\rm eff}$~=~6850~K and log $g$~=~0.2.

\placefigure{T-g}

In this analysis, only lines were used that were of similar strength
(75$-$90~m\AA\ for IRAS~06530$-$0213 and 30$-$40~m\AA\ for HD~56126)
for all species.
In this way the influence of $\xi_{t}$ on the derived $T_{\rm eff}$ and
log $g$ values was minimized, since all of the
lines respond in almost the same way to changes in $\xi_{t}$.
To confirm this, computations were performed with $\xi_{t}$=~3 km s$^{-1}$
and $\xi_{t}$=~7 km s$^{-1}$, and it was found that these resulted in
little or no difference in the values of the intersections,
although the values of the abundances were obviously
changed, by $\pm$0.1 to $\pm$0.2~dex.
The advantage of this method is that one can derive three parameters,
$T_{\rm eff}$, log $g$, and metallicity, simultaneously.
However, to determine more accurate values of these parameters,
one should use neutral and ionized species of many elements,
rather than only two as we were able to do.

\begin{figure}
\figurenum{4}
\caption{Spectroscopic determination of atmospheric parameters
$T_{\rm eff}$, $\xi_{t}$, and log $g$ for IRAS 06530$-$0213 using
Fe~I (x), Fe~II ($\bullet$), and C~I lines (*) for IRAS~06530$-$0213.}
\label{atm-par}
\plotone{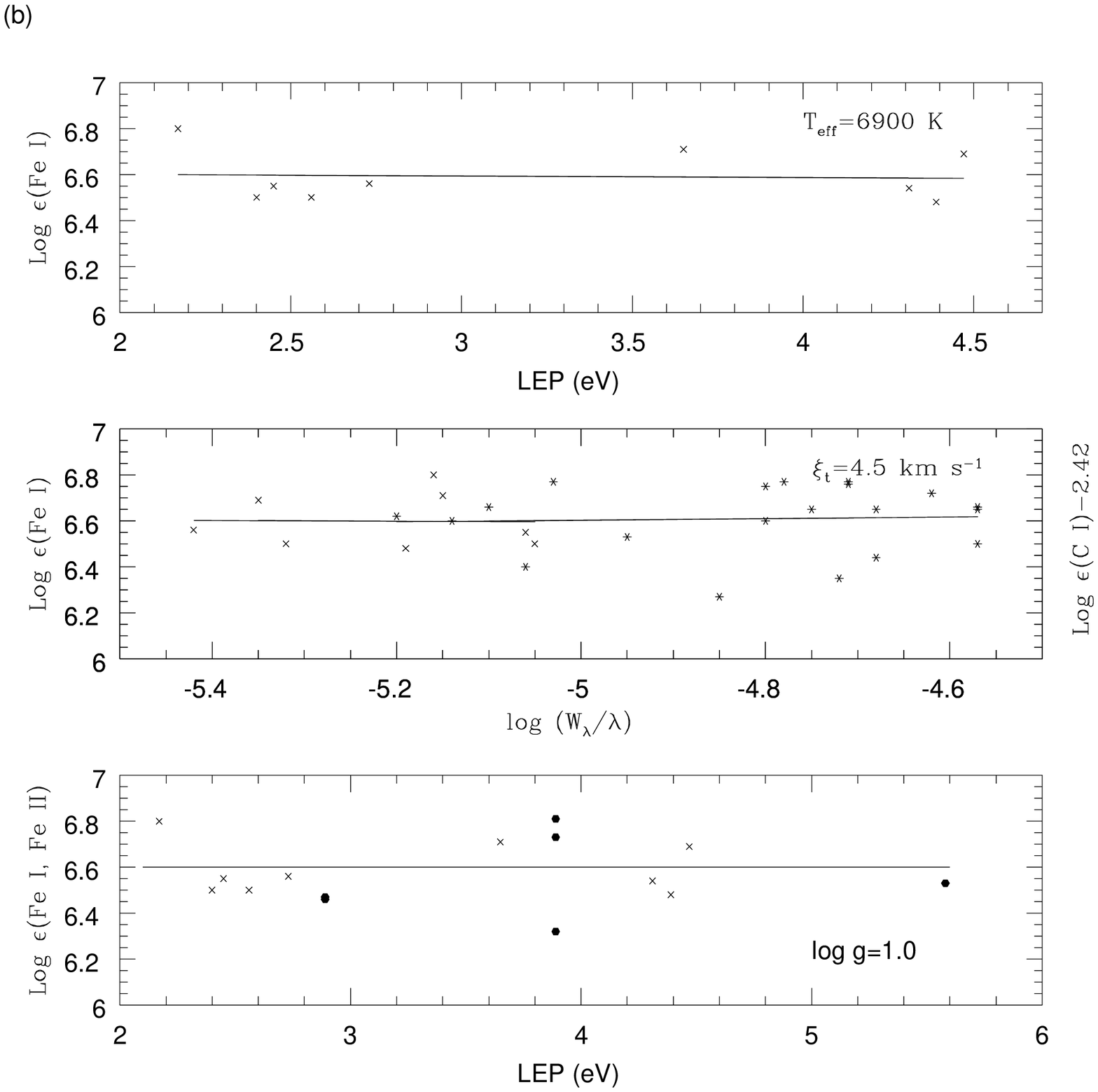}
\end{figure}

As a second approach to determining the atmospheric parameters,
we used the standard procedure of analyzing the excitation and ionization
equilibria of Fe and C\,{\sc i} lines.
The value of $T_{\rm eff}$ was determined by examining the effect of
different $T_{\rm eff}$ values on the derived Fe\,{\sc i} abundances, to find the
value for which the abundances of Fe\,{\sc i} are independent of the
lower excitation potentials (LEPs) of individual Fe\,{\sc i} lines.
To minimize the effect of $\xi_{t}$, weak
(25$-$55~m\AA) Fe\,{\sc i} lines were used.
The value of log $g$ was determined using the ionization balance of
neutral and singly-ionized lines of Fe.
The value of $\xi_{t}$ was determined by comparing abundance versus
reduced equivalent width (W$_{\lambda}/\lambda$) for many Fe\,{\sc i} and 
C\,{\sc i} lines with a wide range of W$_{\lambda}$ (20 - 150~m\AA).
From this, the value of $\xi_{t}$ was determined for which the reduced
equivalent widths of Fe and C were independent of the strength of their
individual lines.
These results are displayed graphically for IRAS~06530$-$0213 in
Figure~\ref{atm-par}.
After many iterations, the following atmospheric parameters were determined:
$T_{eff}$ = 6900$\pm$250~K, log~$g$ = 1.0$\pm$0.5,
$\xi_{t}$ = 4.5$\pm$0.5 km s$^{-1}$ and [M/H] = $-$1.0 for IRAS~06530$-$0213,
and $T_{eff}$ = 7250$\pm$250~K, log~$g$ = 0.50$\pm$0.5,
$\xi_{t}$ = 5.0$\pm$0.5 km s$^{-1}$ and [M/H] = $-$1.0 for HD~56126.
The uncertainties listed are based upon the sensitivities of the abundances
to variations in these parameters.

\placefigure{atm-par}

In this study, the atmospheric parameters derived in three different ways,
(a) low-resolution spectrum, (b) $T_{\rm eff}$-log~$g$ diagram, and
(c) excitation and ionization equilibria, are found to be in good agreement.
The derived atmospheric parameters for HD 56126 in this study are in very
good agreement with those derived for this same star in previous studies
\citep{kloch95, vanwinckel00}.
The adopted model atmosphere parameters, based on our excitation$-$ionization 
analysis, are listed in Table \ref{model-atm} and are used in the subsequent 
abundance analyses.

\placetable{model-atm}

\subsection{Abundances}
\label{abundance}

In deriving the elemental abundances, the widely used plane-parallel,
line-blanketed, local thermodynamic equilibrium (LTE) stellar model
atmospheric grids of Kurucz were used, which are computed using the
ATLAS9 code\footnote{See the Kurucz web site at http://cfaku5.harvard.edu.}.

Abundances were primarily computed through the fine analysis method,
in which the theoretical strength of each line is computed and
the model abundances are iterated until the computed line strengths are
matched with the observed ones.
The fine analysis was done using a modified version of the computer
program MOOG \citep{sneden73}.
The results of this analysis are summarized in Table~\ref{abund}.
In the table, $n$ represents the number of lines measured,
log~$\epsilon$(X) is the logarithmic abundance of the particular element (X)
with respect to a hydrogen abundance of log~$\epsilon$(H)~=~12.0,
and [X/H] and [X/Fe] are the logarithmic ratios with respect to hydrogen and
iron, respectively, relative to the solar values.
The solar values of \citet{grevesse98} have been used except for C and O,
for which we have adopted more recently derived solar abundance values
\citep[log $\epsilon$(C)=8.39, log $\epsilon$(O)=8.69;][]{allende02}.
The line-to-line scatter is indicated by the standard deviation $\sigma$.

\placetable{abund}

\subsubsection{Carbon, Nitrogen, and Oxygen}

\begin{figure}
\figurenum{5}
\caption{Spectral synthesis of the wavelength regions 7115 \AA~ and 6156 \AA~
for IRAS 06530$-$0213.
The spectra computed from the adopted model fits well with the
observed spectra (points) for the abundances given in Table~3.}
\label{CO-spec}
\plotone{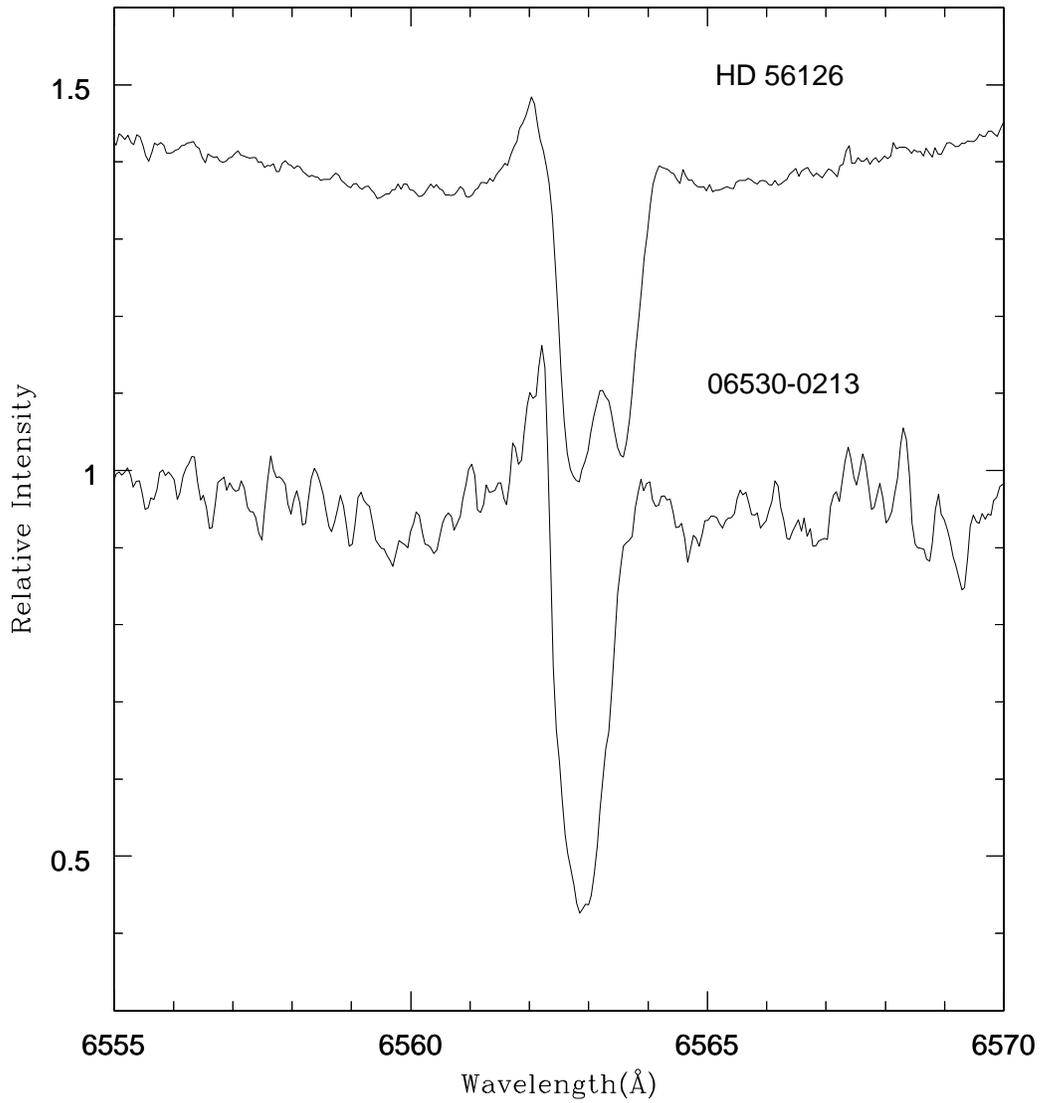}
\end{figure}

Abundances of C and N were derived from a large number of lines.
Both stars are found to be enhanced in C and N relative to Fe,
[C/Fe] = 1.1 and [N/Fe] = 1.0 for IRAS~06530$-$0213 and
[C/Fe] = 0.7 and [N/Fe] = 0.9 for HD 56126.
The oxygen abundances were determined from analyses of the 
O\,{\sc i} triplet at 6156~\AA, and abundances of 
[O/Fe] = 0.5 (IRAS~06530$-$0213) and [O/Fe] = 0.7 (HD~56126) were derived.
Thus both stars are overabundant in O relative to Fe.
Comparison of the computed spectrum with the observed
spectrum in the region of O\,{\sc i} triplet and in the region of the C lines at
7115\,\AA\ is shown for IRAS~06530$-$0213 in Figure \ref{CO-spec}.

\placefigure{CO-spec}

Recent studies of C and N  suggest that there are significant non-LTE
effects in this temperature and low-gravity domain and in low-metallicity stars.
The study of non-LTE effects on the neutral C lines in the atmospheres
of F- and A-type dwarfs \citep{rentz96} suggests that non-LTE effects
are significant even for dwarfs.
\citet{venn95} estimates a similar non-LTE correction for the F-type supergiant
HR 6130.
The non-LTE studies of N\,{\sc i} lines by \citet{luck85} and \citet{venn95}
show that N\,{\sc i} lines are significantly affected by non-LTE effects.
However, the corrections for our two stars may be even slightly higher,
since both HD~56126 and IRAS~06530$-$0213 are more metal-deficient than
the stars considered in the above studies.  Non-LTE effects become more
significant in metal-poor stars due to over-ionization of neutral elements
as a result of the stronger UV radiation field.
These studies suggest non-LTE corrections of approximately $-$0.3 dex and
$-$0.4 dex for the abundances of C and N, respectively.
We note these possible corrections to the abundances derived for C and N,
but have not applied them to the abundance values listed in Table~\ref{abund}.
In the case of the O\,{\sc i} abundance determined from the 6156~\AA\ triplet,
the deviations from LTE were found to be not significant ($\leq$ 0.1~dex) in stars similar
to those in this study \citep{takeda98}.

\subsubsection{Metals}

The metallicity was determined using neutral and ionized iron-peak lines
of W$_{\lambda}\leq$ 100 m\AA.  Both of these stars were found to be
metal-poor, with [Fe/H]=$-$0.9 for IRAS~06530$-$0213 and [Fe/H]=$-$1.0 for
HD~56126.
The abundances of other species like Ca, Sc, Ti and Cr
support the metal-deficiency in these two stars.
The recent non-LTE study of \citet{rentz96} suggests a positive
non-LTE correction to the LTE Fe abundance.
For a group of A-F stars, that study revealed that non-LTE effects are
significant for metal-poor and low-gravity stars as a result of
over-photoionization and low collisional rates, respectively;
it indicates a non-LTE correction of approximately $+$0.1 dex
for a star having physical parameters similar to the stars in our study.
This is within the uncertainties of the Fe abundance
and no correction was applied.

\subsubsection{s-process Elements}

The abundances were determined for several s-process elements:
Y, Zr, Ba, La, Ce, Nd, Sm, Eu, and Dy for HD~56126 and
Y, Ba, La, Ce, Pr, and Nd for IRAS~06530$-$0213.
In the case of HD~56126, the abundance results are based on many good,
weak lines for each species, but this was not possible in the case of
IRAS~06530$-$0213, since we were not able to find as many weaker lines
(W$_{\lambda}$$<$150~m\AA).
This limitation for IRAS~06530$-$0213 seems to be due partly to its larger
enhancement in the heavy elements (see Fig. \ref{high-resS}) 
and partly to its more limited spectral range in the blue.
Both stars are very abundant in s-process elements as compared with
iron, with average values of [s-process/Fe]~=~1.9 for IRAS~06530$-$0213
and [s-process/Fe]~=~1.6 for HD~56126.
Note that very strong Ba lines ($W_{\lambda}$ $\geq$ 450\,m\AA) are present 
in the spectra of IRAS~06530$-$0213 and the
derived abundances might be an overestimate (Table~\ref{abund}).
Thus we did not include the Ba abundance in computing the
s-process average for this star.

\subsubsection{Comparison with Recent Study of IRAS~06530$-$0213}

An abundance analysis of IRAS~06530$-$0213 has recently been carried
out by \citet{rey02}.  This is based upon high-resolution, high S/N 
spectra obtained with the ESO 8-m VLT.  Our atmospheric models are
in good agreement except for $T_{\rm eff}$, for which our model is cooler  
by 350 K; this results in a significantly lower metalicity ([Fe/H]=-0.9 versus 
-0.5 for Reyniers).  However, the abundances relative to Fe ([X/Fe]) are
in good agreeement.  Thus he similarly finds the object to be carbon-rich and
overabundant in s-process elements.

\subsubsection{Comparison with Previous Studies of HD~56126}

HD~56126 has been previously analyzed by \citet{partha92}, \citet{kloch95},
and more recently \citet{vanwinckel00}, and  the present
results can be compared with the results of these studies.
Klochkova, with spectra of R $\simeq$ 24,000 in the red,
determined atmospheric parameters
similar to ours; she found $T_{\rm eff}$~=~7000~K, log~$g$~=~0.1, and
$\xi_{\rm t}$~=~5.5~km s$^{-1}$.
Our results for the abundances of C, N, O, Sc, and Fe are in general
agreement with the values of Klochkova.
However, we find the abundances of Mg, Si, Ca, Ti, and Cr to be significantly
less ($>$~0.5~dex) than the values she derived, with our values being more
consistent with the low-metallicity of HD~56126.
We also obtain similar values to Klochkova for the abundances of the
ionized species of the s-process elements (Y, La, Nd).  
The recent study of \citet{vanwinckel00} is based upon spectra with a similar
resolution as in our study and includes an even larger spectral range resulting
in the measurement of an even larger number of good lines (343 compared with
207 in our study).  Their atmospheric model is identical to ours;
they find $T_{\rm eff}$~=~7250~K, log~$g$~=~0.5, $\xi_{\rm t}$~=~5.0~km s$^{-1}$,
and [M/Fe]~=~$-$1.0.
Our results are in good agreement with theirs; they obtain
[C/Fe]~=~+1.08, [N/Fe]~=~+0.85, and [O/Fe]~=~+0.81.
For the cases in which our results differ from those of Klochkova, they are
in agreement with those of Van Winckel \& Reyniers.
Our results show excellent agreement with Van Winckel \& Reyniers in the
high abundance of s-process elements, with an average difference of only
+0.12 dex for the six such elements we have measured in common 
(for which more than one line was measured).
Parthasarathy et al. only obtained spectra around eight particular wavelength
settings in the red, with a range of 50$-$60 \AA\ each, and thus their
spectral coverage is small.  Their spectra have a resolution of
$\sim$55,000 at H$\alpha$.  They ran a series of models, including one
close to ours: $T_{\rm eff}$~=~7000~K, log~$g$~=~0.5, and
$\xi_{\rm t}$~=~4~km s$^{-1}$.  
Our results for N and O are in good agreement with their values, 
but our abundances of C and Ca differ from theirs by $\sim$~$-$0.5~dex, 
and we do not find the large S overabundance that they claim 
(based upon only one line).
In summary, our results are in good agreement with those of Van Winckel \&
Reyniers but differ in important ways from the lower-resolution study of
Klochkova and the limited spectral study of Parthasarathy et al.
Hereafter, we discuss only our current results for this star.

\subsection{Uncertainties}

The uncertainties in the derived abundances due to the line-to-line scatter,
$\sigma$, are attributed to the uncertainties in the measurements of the 
W$_{\lambda}$ and to the accuracy of the {\it gf} values.  
These are listed in Table \ref{abund}, and the consequent uncertainty in the 
abundance value is $\sigma$/$\sqrt{n}$.
For the abundances derived from fewer than three lines of a
species, we assume a realistic uncertainty to be 0.20~dex.

The uncertainties in the abundance values due to uncertain model parameters
were computed for both stars. The uncertainty in  $\xi_{t}$
($\pm$0.5~km s$^{-1}$) has little effect on the abundances ($<$0.05~dex).
This is largely due to our use of weak lines in the analysis.
The uncertainty in log~{\it g} ($\pm$0.5~dex) has a larger effect,
typically $\leq$$\pm$0.1 but reaching around $\pm$0.2 in a few
cases (N\,{\sc i}, Na\,{\sc i}, Mg\,{\sc i}, Ca\,{\sc i}, Fe\,{\sc i}).
The uncertainty in $T_{\rm eff}$ ($\pm$250 K) has the largest effect,
typically $\pm$0.1 to $\pm$0.2 dex for most of the elements and
around $\pm$0.3 dex for the $s$-process elements.

\subsection{H${\alpha}$, Na, and K Profiles}

\begin{figure}
\figurenum{6}
\caption{ The H$\alpha$ profiles from IRAS~06530$-$0213 and HD~56126,
showing shell-like emission structure and multiple components within
the absorption feature.}
\label{Halpha}
\plotone{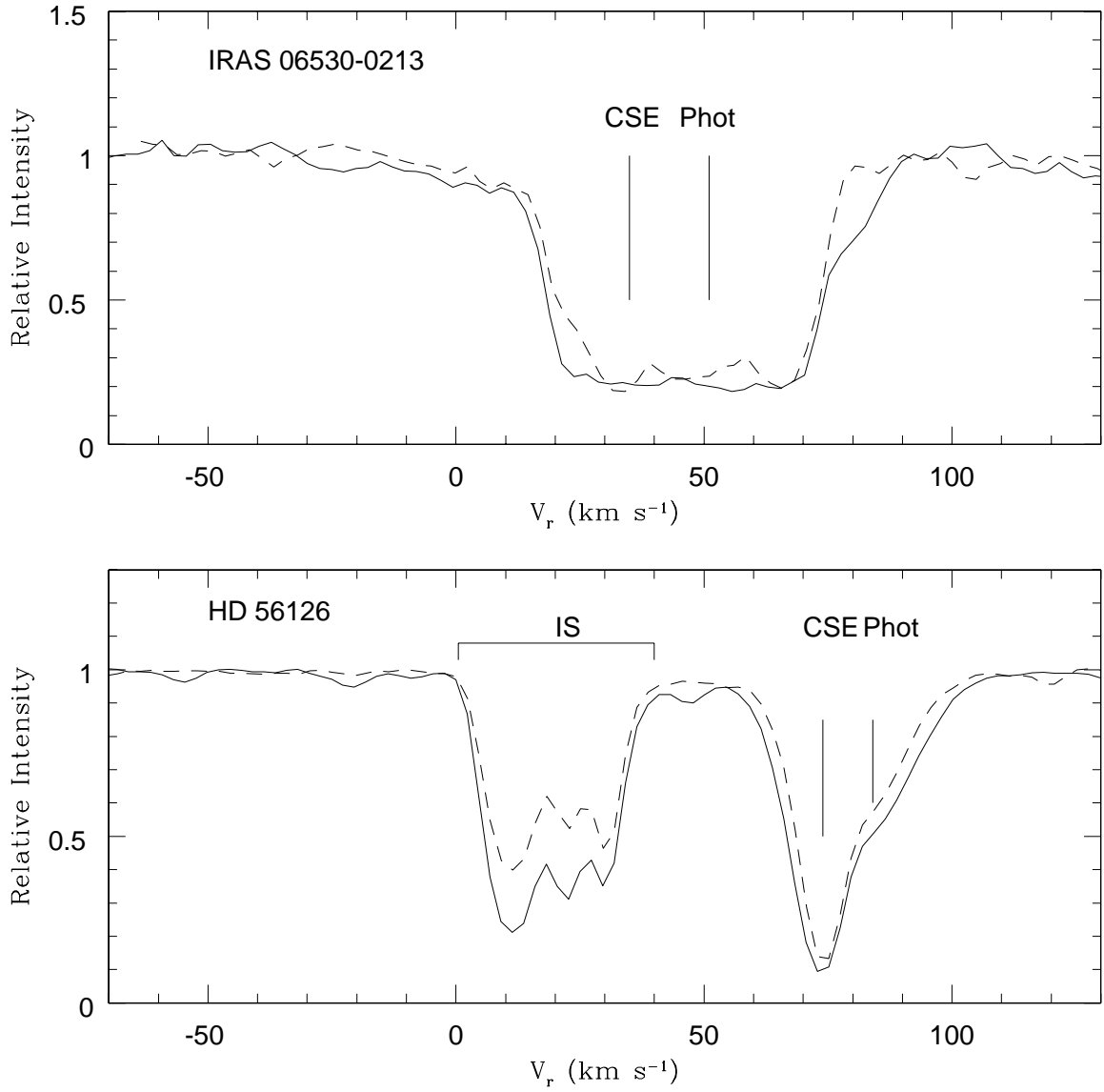}
\end{figure}

The strongest features in both spectra are the H${\alpha}$ and Na D$_{1}$
and D$_{2}$ lines.  The H${\alpha}$ lines display shell-like
emission profiles with a central absorption and are shown in Figure
\ref{Halpha}.
This sort of H${\alpha}$ profiles, a central absorption within the emission,
we have found to be common in PPNs of F spectral type.

\placefigure{Halpha}

Previous studies of the H${\alpha}$ profile in HD~56126 have been published
\citep{oud94, lebre96, barthes00}.
They all show a shell-like profile which is found to vary over a time scale
of weeks.  While an indication of two absorption components is seen in
many of the published profiles, in our new spectrum of 1997 October 17
they are more clearly resolved.   
A difference of 34.9 km~s$^{-1}$ is measured for the two components.  
The spectrum of 2002 November 14 shows instead a typical shell spectrum,
with emission peaks of approximately equal heights on both sides of a
deep absorption feature.
Several studies have attributed this variable
H${\alpha}$ emission to atmospheric pulsations and the propagation
of associated shock waves through the photosphere, which
generate emission with some self-absorbed \citep{lebre96, jeannin96}.
Pulsations have been deduced from the observations of light and
velocity variability, with a period of ~37 d cited by \citet{barthes00}.
For IRAS~06530$-$0213, the two spectra obtained at different times 
appear very similar.

\begin{figure}
\figurenum{7}
\caption{The complex Na~I D$_{2}$ and D$_{1}$ absorption profiles
from IRAS~06530$-$0213 and HD~56126 are plotted versus
radial velocity.  The velocity of the photospheric (Phot) and
circumstellar (CSE) components are marked.
For HD~56126,  interstellar (IS), circumstellar, and stellar
profiles are well separated. However, for IRAS~06530$-$0213 this is not
the case, with both stellar and circumstellar components embedded
among the interstellar components.}
\label{Na-D}
\plotone{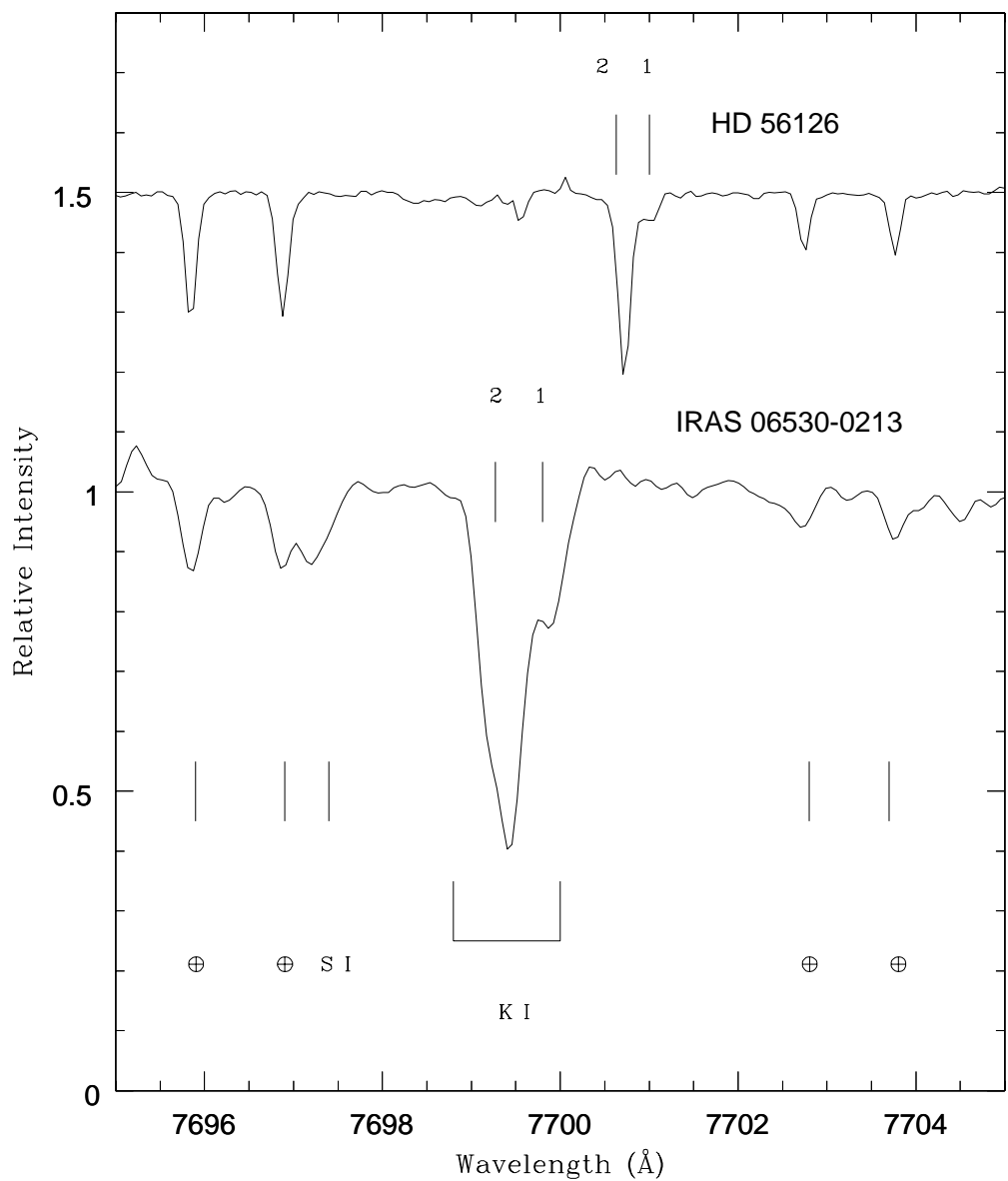}
\end{figure}

The Na~D profiles are very strong in the spectra of IRAS~06530$-$0213 and
HD~56126, reaching depths of 0.1 compared to the normalized continuum.
This spectral regions is displayed for both stars in Figure \ref{Na-D}, with
the Na D$_{1}$ and D$_{2}$ line profiles superimposed to show their similarities.
The expected positions of the photospheric and circumstellar components
are indicated in the figure, based upon the derived velocities of atomic and
molecular (Sect. 4.7) absorption lines, respectively.
For IRAS~06530$-$0213, the nearly constant strengths of these very broad
profiles suggest that they are all saturated.
Analysis of the D$_{1}$ and D$_{2}$ profiles for this object leads to the
suggestion that there may be as many as six different absorption components
with overlapping profiles, which makes it difficult to derive quantitative results.
The photospheric and circumstellar absorption lines appear to be embedded in
a forest of interstellar Na lines.
These are likely to be due to foreground interstellar clouds, given the low
galactic latitude ($b$=-0.1$^o$) of IRAS~06530$-$0213, and their velocity
differences are too small to be resolved in our spectra.
For HD~56126, the photospheric absorption component and a strong
circumstellar component are seen, along with three interstellar components
(V$_{\rm r}$ = 12, 23, 30 km~s$^{-1}$).
Similar results were found by \citet{bakker96}.
The circumstellar component has a velocity of 74.0 km~s$^{-1}$.

\placefigure{Na-D}

\begin{figure}
\figurenum{8}
\caption{The K~I absorption profiles at 7699~\AA.
Components 1 and 2 in both profiles arise in the photosphere and in the
circumstellar envelope, respectively.}
\label{Kabs}
\plotone{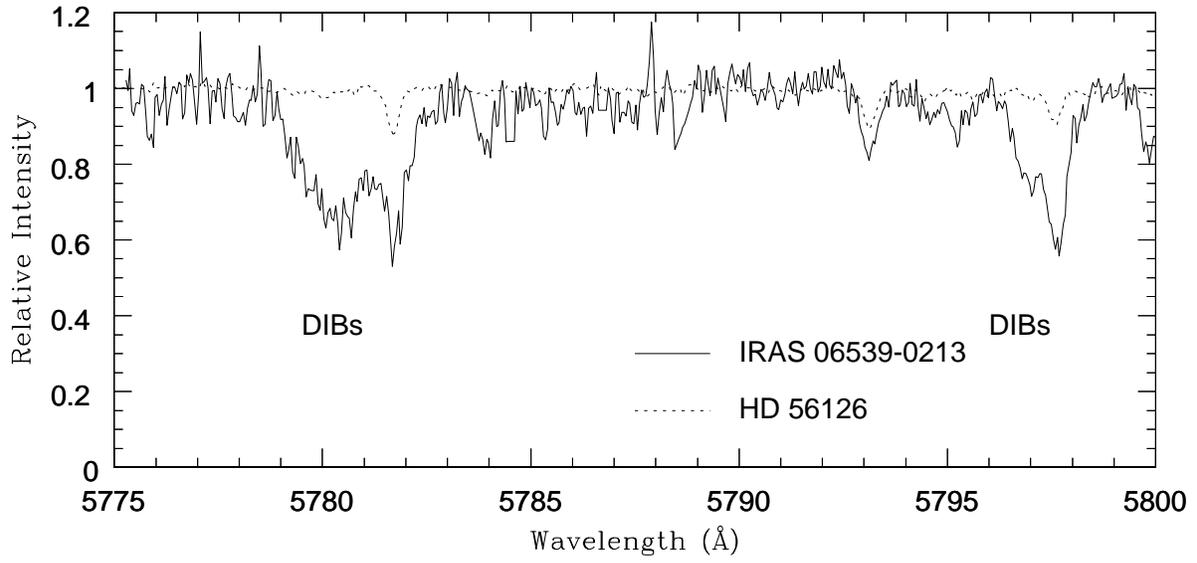}
\end{figure}

The profiles of neutral K at 7700 \AA\ for both the stars are shown in
Figure \ref{Kabs}. Each profile shows a strong and a weak component.
For each star, the weaker component agrees with the radial velocity of the
photosphere and the stronger component agrees with the velocity of the
cooler circumstellar envelope (see Table\,5).
The stronger component in IRAS~06530$-$0213 is slightly asymmetric
and appears be blended with another component shifted 
10 km~s$^{-1}$ blueward.
Ultra-high-resolution spectral observations of HD~56126 by \citet{crawford00}
have resolved this circumstellar K\,{\sc i} line into two (or perhaps three) 
components with a velocity difference of 1.0 km~s$^{-1}$; 
they interpret this as due to discrete shells in the CSE.

\placefigure{Kabs}

\begin{figure}
\figurenum{9}
\caption{Diffuse interstellar bands are seen in the spectrum of
IRAS~06530$-$0213 (solid line) but not in the spectrum of HD~56126
(broken line).}
\label{dibs}
\plotone{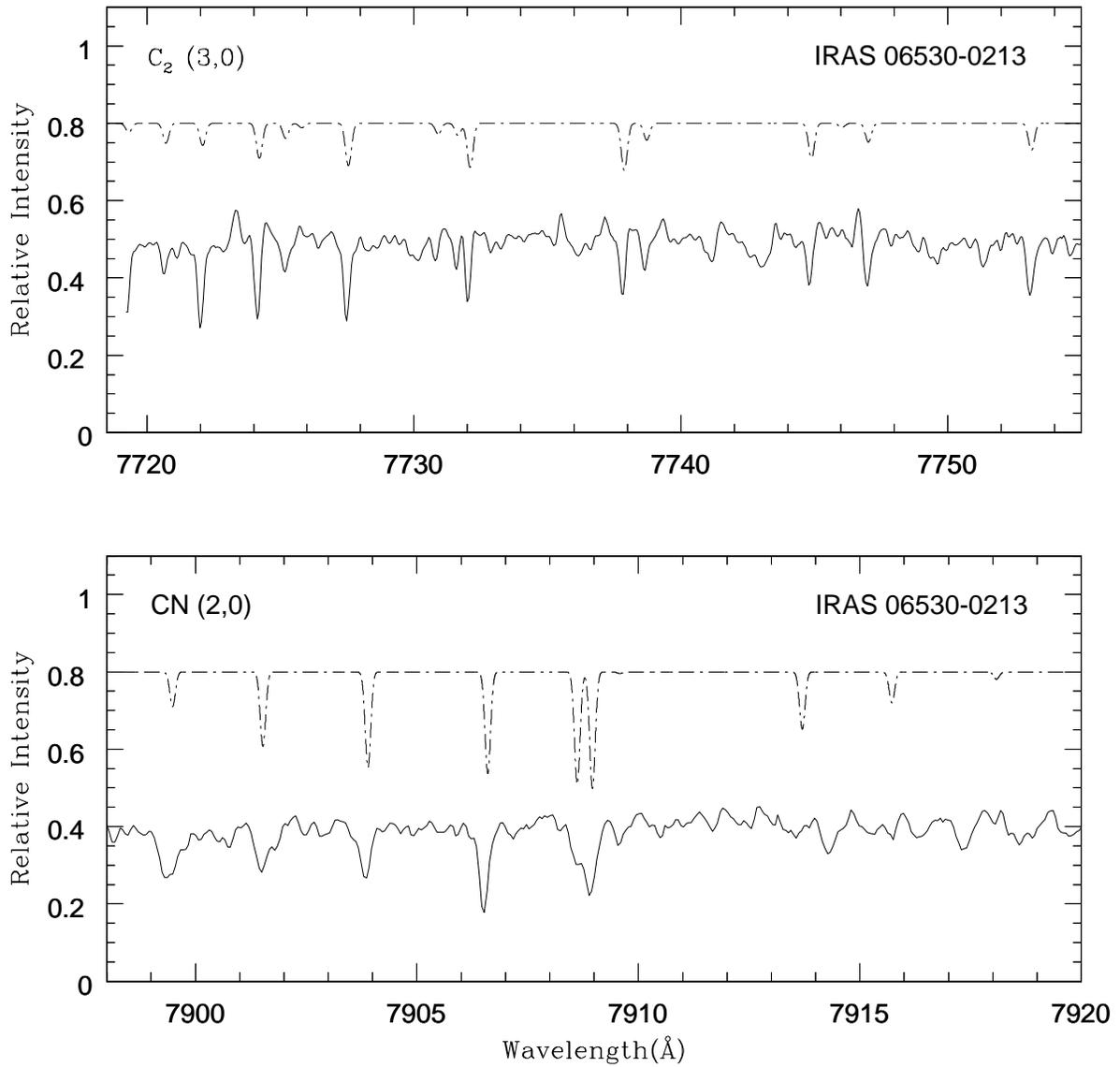}
\end{figure}

We also searched for diffuse interstellar bands (DIBs) in the spectra of these two stars.
No DIBs were found in the spectra of HD~56126.
However, strong DIB features were identified in the spectra of IRAS~06530$-$0213
at 5780.41~\AA, 5797.0~\AA, 6376.1~\AA, and 6379.3~\AA.
Some of these are shown in Figure \ref{dibs}.
The radial velocities of the DIBs are similar to those of the photospheric lines
and differ from those of the circumstellar molecular lines.
These results differ from the those of the four PPNs studied by \citet{kloch01};
they concluded that DIBs formed (mainly) in the circumstellar envelopes.
Since the DIBs are not expected to form in the warm photospheres of these stars,
we interpret the results for IRAS~06350$-$0213 to imply that they formed in an
interstellar cloud(s) with a velocity close to the photospheric velocity.
The broad, blended Na~D profiles are consistent with the idea that such
interstellar clouds are present along the line of sight with velocities similar to
that of the photosphere.

\placefigure{dibs}

\subsection{Molecular Lines}

\begin{figure}
\figurenum{10}
\caption{C$_{2}$ Phillips system band and CN Red system
band absorption features in the spectrum of IRAS~06530$-$0213 
(solid line), along with sample synthetic spectra (dash-dot line).}
\label{moles}
\plotone{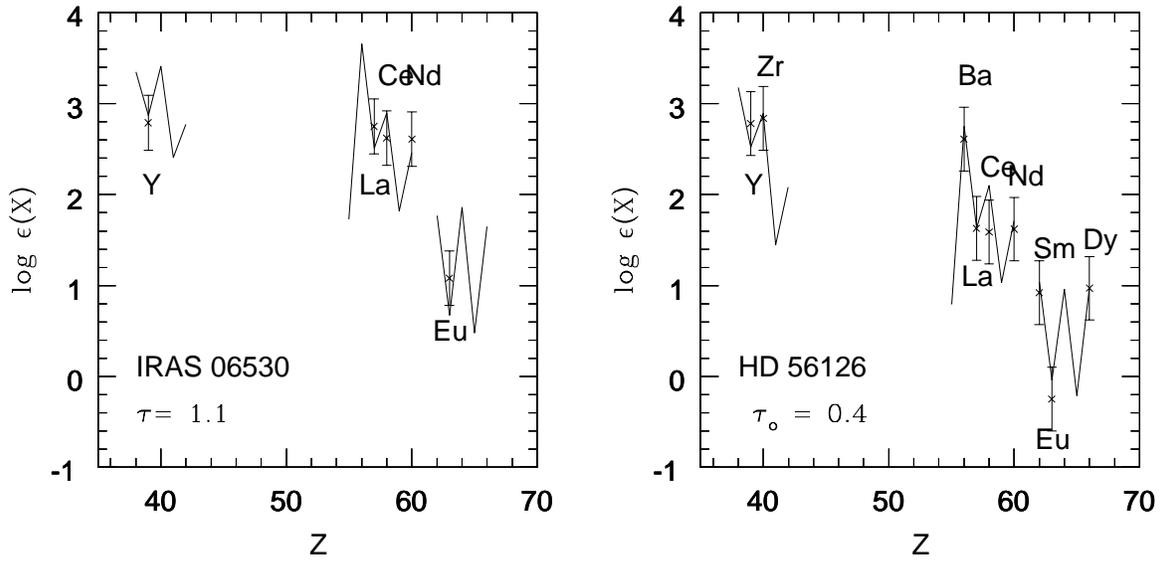}
\end{figure}

Molecular absorption bands of C$_{2}$ Phillips (3,0) and CN Red System (2,0)
were detected in the red part of the visible spectra of both IRAS~06530$-$0213
and HD~56126.
They were identified using the molecular band wavelengths listed by
\citet{bakker95}.
These bands in HD~56126 have previously been studied by \citet{bakker96}.
Sample spectra displaying molecular bands in IRAS~06530$-$0213
are shown in Figure \ref{moles}. The C$_{2}$ and CN bands are red-shifted
by 36.0$\pm0.6$ and 37.4$\pm1.5$ km s$^{-1}$, respectively, relative to their
rest wavelengths. Thus these lines are blue-shifted by $\sim$14 km s$^{-1}$
with respect to the atomic lines, which are of photospheric origin.
Such a shift is expected if these bands are produced in a cool, expanding
circumstellar envelope.
Since the photospheres of these transition objects are known to pulsate
with amplitudes of several km s$^{-1}$ \citep{hrivnaklu99, barthes00},
we will assume that the previously measured CO velocity is the system velocity.
This results in an envelope expansion velocity of V$_{exp}$~=~13 km s$^{-1}$.
The velocities were similarly measured for these bands in HD~56126. The measured
CO velocity $V_{\rm r}$ = 87.5 km s$^{-1}$
($V_{\rm lrs}$ = 73 km s$^{-1}$; Omont et al. 1993) and our measured CSE molecular lines would result in
V$_{exp}$ = 10 km s$^{-1}$ for HD~56126.
The measured expansion velocities are summarized in Table \ref{vel},
and they are typical of the expansion velocities found for PPNs.
(For other PPN expansion velocities based upon visible spectra, see
\citet{bakker97} and  \citet{reddy99}.)

\placefigure{moles}

\placetable{vel}

\section{DISCUSSION}

As part of an ongoing, multi-wavelength effort to identify and study
PPNs, we have analyzed our observations of IRAS~06530$-$0213.
The object is shown to be similar spectroscopically to the well-studied,
carbon-rich PPN HD~56126 and to possess various characteristics of a PPN.
These include a double-peaked SED, an expanding CSE, and the general
abundance pattern of a post-AGB object;
we will elaborate on these below.
In addition, a small (2$\farcs$4 $\times$ 1$\farcs$2) bipolar reflection nebula
is seen around IRAS~06530$-$0213 in visible light \citep{ueta00}.

\subsection{Spectral Energy Distribution}

\begin{figure}
\figurenum{11}
\caption{Spectral energy distribution for IRAS~06530$-$0213,
showing typical double-peaked distribution seen from PPNs.
The peak in the visible$-$near-infrared is from the reddened
photosphere ($T_{\rm fit}$$\sim$2500 K) and the peak in the
mid-infrared arises from the emission of the circumstellar
dust (T$_d$$\sim$170 K).  The upper dashed curve shows the
estimated correction for interstellar extinction.}
\label{sed}
\plotone{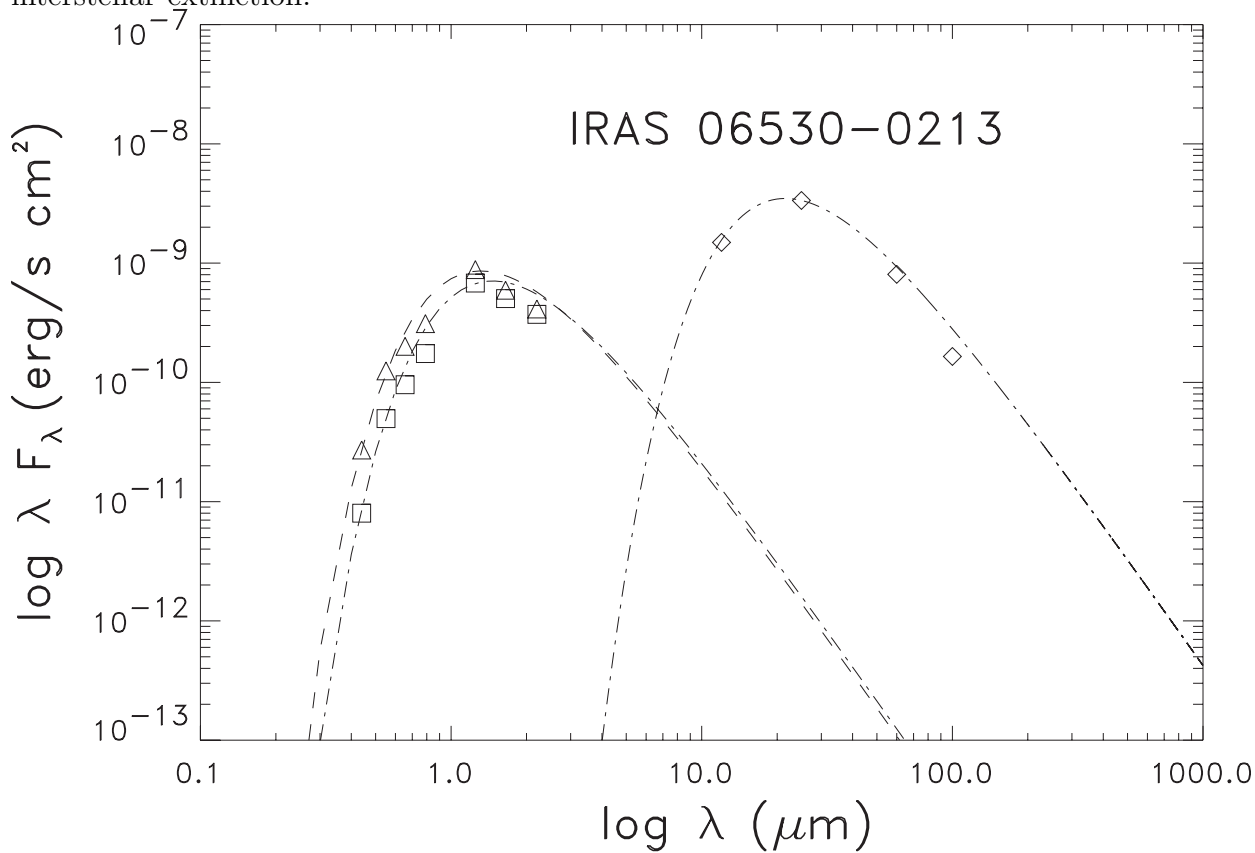}
\end{figure}

Combining these new visible and near-infrared photometry with the
{\it IRAS} flux measurements, the SED of IRAS~06530$-$0213 can be
delineated.  This is shown in Figure \ref{sed}.  One can clearly see the
double-peaked shape of the SED, with one peak arising from the
reddened photosphere and a larger one from the circumstellar dust.
Since the object is in the galactic plane (l=215$^{o}$, b=$-$0.1$^{o}$),
some of the reddening and extinction will be caused by interstellar dust.
Also shown is the SED corrected for this.
An estimated value of A$_V$ $\approx$ 1.0 mag was determined from the
interstellar extinction study of \citet{neckel80} and the interstellar
extinction law of \citet{cardelli89} was applied.
The fitted (reddened) photospheric temperature is $T_{\rm fit}$ = 2500~K, rising to
T$_{\rm fit}$$^\prime$= 2800~K when corrected for interstellar extinction;
thus significant circumstellar reddening exists.
The dust temperature is $T_{\rm d}$ = 170~K.
The total flux from the dust component is $\sim5$ $\times$ 10$^{-9}$
ergs cm$^{-2}$ s$^{-1}$, while the flux from the photosphere is
one-fifth this amount.  
These values are based upon black body fits to the SED. 
These lead to an observed luminosity of $\sim$180 (D/kpc)$^2$ L$_{\sun}$.
The SED of HD~56126 is similar in appearance, although there the two
peaks are of similar height \citep{hrivnak89}.
A detailed modeling of the SED of HD~56126 found $T_{\rm d}$ = 165~K and an
observed luminosity of $\sim$1400 (D/kpc)$^2$ L$_{\sun}$ \citep{hrivnak00}.
Thus if the two objects have the same intrinsic luminosity,
IRAS~06530$-$0213 is 2.8 times as distant.

\placefigure{sed}

\subsection{Circumstellar Envelope}

The CSE of IRAS~06530$-$0213 had previously been detected in
CO (2$-$1) millimeter-line emission at 1.3 mm \citep{hu94}.
In these new high-resolution spectroscopic observations, the CSE is
detected by its absorption of stellar light in the molecular lines of
C$_2$ and CN and in an atomic line of K\,{\sc i}.
The expansion velocities determined from these are in the range 13$-$16
km s$^{-1}$;
these are slightly larger than the cited expansion velocity of 10 km s$^{-1}$
determined from the CO millimeter-line observations.
However, this latter value is measured as half of the total width of the CO
emission feature, and visual inspection of the published CO spectrum shows
that it is noisy and may well be larger than 10 km s$^{-1}$.
The expansion velocity for HD~56126 is slightly lower, $V_{\rm exp}$ = 10$-$13
km~s$^{-1}$ as measured from the new C$_2$, CN, K\,{\sc i}, and Na\,{\sc i}
spectral observations; this is in very good agreement with
$V_{\rm exp}$ = 10.2 km~s$^{-1}$ measured from the millimeter-line CO
observations \citep{omont93}.

\subsection{General Abundance Patterns}

The abundances of CNO elements are crucial in determining the stellar
evolutionary phase and the internal mixing and nucleosynthesis processes.
The enhancement of the CNO elements in the photospheres of HD~56126
and IRAS~06530$-$0213 is a clear indication that these stars are evolved.
Both the stars are found to have excess total CNO abundance 
($\Sigma$CNO$_{\rm obs}$ = 8.9 for IRAS~06530$-$0213,  8.7 for HD\,56126) 
relative to the expected value ($\Sigma$CNO$_{\rm exp}$ = 8.20). 
The expected value was determined by assuming initial values of 
[C/Fe]$\simeq$ 0.2 \citep{gusta99}, [O/Fe] $\simeq$ 0.40 \citep{nissen02}, 
and [N/Fe] = 0.0.
The 1st dredge-up in the RGB phase and the 2nd dredge-up in the AGB 
phase alter the C, N, and perhaps O abundances while keeping the total 
CNO abundance unchanged.
An excess total CNO abundance indicates that a star has experienced the 3rd 
dredge-up, which mixes freshly produced He-burning products into the 
photosphere.  An examination of the individual C, N, and O abundances
in Table\,3 shows that all three elements are significantly
overabundant relative to iron. 
An enhancement of C is expected as a result of the burning of He via the
triple-$\alpha$ reaction and the deep convection during the AGB phase, 
and an enhancement of N is expected due to the CN reaction during both the
RGB and AGB phases. 
Interestingly, O is overabundant in both stars, especially in HD\,56126. 
In the case of IRAS~06530$-$0213, the value of [O/Fe] = 0.5 is consistent 
with the chemical evolution of the Galaxy.
The value of [O/Fe] = 0.7 in HD\,56126 is 0.3~dex in excess of the expected
initial O abundance. 
Since it is the ratio of abundances, an uncertainty in the [O/Fe] ratio as a 
result of uncertainties in the model parameters is small ($\pm$0.1~dex) 
and non-LTE effects are small ($\le$0.01; Sec. 4.4.1). 
It is possible that the O overabundance in HD\,56126 is the result of AGB mixing.
The He triple-$\alpha$ in the interiors of AGB stars results in the production of
additional C and the additional O may be produced as a result of 
the $^{12}$C($\alpha$, $\gamma$)$^{16}$O reaction. 
While standard convective mixing models do not predict an overabundance of 
O in the photosphere, mixing models with convective overshoot
do predict a significant enhancement of both  C and O \citep{herwig97}.
Thus the overabundances of both C and N indicate that IRAS~06530$-$0213
has evolved through the RGB and AGB phases of evolution similar to HD~56126, 
and the overabundance of O is consistent with this.

Among $\alpha$-process elements, abundances of Mg, Si, Ti, and Ca were 
measured for IRAS~06530$-$0213 (Mg, Si, S, Ca, and Ti for HD~56126),
and these are found to be normal relative to Fe for stars of [Fe/H] = $-$1.0.
Slight overabundances of some of these elements are consistent with 
the Galactic chemical evolution \citep[see][]{chen02}.

The enhancement of heavy elements like Y, Zr, Ba, La, etc. is one of the distinctive
features of the photospheres of AGB/post-AGB stars.
These elements are produced by iron-seed nuclei capturing neutrons slowly
relative to beta-decay (the ``s-process'').
In low-mass AGB stars neutrons are mainly produced from
the reaction $^{13}C(\alpha,n)^{16}O$ \citep{iben83}.
This is accomplished by the diffusion of protons downwards into the He shell
where they are captured by the abundant $^{12}$C,
producing a sufficient amount of $^{13}$C.
The heavy elements produced in the interior are brought to the photosphere
through the third dredge-up during the thermally-pulsing AGB phase.

The particular importance of measuring the abundance of the s-process
elements is that this can be used to determine a star's particular phase of
evolution and also to better understand the nucleosynthesis processes
(from the s-process elemental distribution).
The large enhancements in the observed s-process elements in
IRAS~06530$-$0213 ([s-process/Fe]=1.9) and HD~56126 ([s-process/Fe]=1.6)
fall in the range of values of 0.9 to 2.2 measured in other PPNs
\citep{reddy99, vanwinckel00, reddy02}.
This is consistent with the other evidence that these two
F-type supergiants are indeed post-AGB stars and have experienced
s-process nucleosynthesis and deep convective mixing on the AGB phase of
evolution.

To better interpret the observed s-process elements,
following the definition of \citet{luck91}, one can derive the ratio of heavy-to-light
s-process elements, [hs/ls] = [hs/Fe] $-$ [ls/Fe], where [hs/Fe] and [ls/Fe] are the
average heavy (La, Ce, Nd) and light (Y, Zr)
s-process elemental abundances, respectively, relative to Fe.
This ratio represents the s-process distribution;
the higher this ratio, the higher the value of the neutron exposure $\tau_{o}$.
The values of $\tau_{o}$ were derived using the theoretical computations
of \citet{busso95} for exponential distribution models of neutron density
$N_{n}$ = 10$^{8}$ cm$^{-3}$.
For HD~56126, [hs/ls]~=~$+$0.1 which corresponds to $\tau_{o}$~$\simeq$~0.4, 
while for IRAS~06530$-$0213, [hs/ls]~=~0.5 which
suggests a neutron exposure rate of $\tau_{o}$ $\approx$ 1.0.

The value of [hs/ls] = 0.5 for IRAS~06530$-$0213 is among the largest found for a
PPN and is similar to the result for IRAS~22272$+$5435 and IRAS\,05113+1347
\citep{reddy02}.
With its associated large overabundance of s-process elements,
IRAS~06530$-$0213 follows the tight correlation between
these two parameters found for six other C-rich PPNs by \citet{vanwinckel00}.
This PPN also follows the correlation of increasing [hs/ls] with decreasing
metallicity
[Fe/H] \citep{reddy02} and fits closely to the predicted theoretical curve (Busso et
al. 2001).
Thus IRAS~06530$-$0213 fits well with the s-process distribution and metallicity
relations found for other PPNs.

\section{CONCLUSIONS}

Atmospheric parameters and elemental abundances of C, N, O, iron-peak,
and s-process elements were derived for IRAS~06530$-$0213 based on
high-resolution spectra and LTE model atmosphere analysis.
Model atmospheric parameters T$_{eff}$ = 6900~K, log~$g$ = 1.0, and
$\xi_{t}$ = 4.5 km s$^{-1}$ were determined.
The star is metal-poor ([Fe/H]=$-$0.9), carbon-rich (C/O=2.6), and enhanced in
C ([C/Fe]=1.3), N ([N/Fe]=1.0), and s-process elements ([s-process/Fe]=1.9).
These abundance results imply that IRAS~0653$-$0213 is a low-mass,
carbon-rich, post-AGB star that has passed through the third dredge-up on
the AGB.  It is similar to the well-studied, C-rich PPN HD~56126.

Circumstellar molecular (C$_2$ and CN) and atomic (K~I) features were
measured that reveal the gaseous component of the expanding CSE.
System and circumstellar expansion velocities of V$_{r}$ = 51 km s$^{-1}$
and V$_{exp}$ = 14 km s$^{-1}$, respectively, were found.
These are in general agreement with previous millimeter-wave CO
measurements.

IRAS~06530$-$0213 can thus be added to the small group of approximately
a dozen identified carbon-rich PPNs.  Like them, it is metal-poor with a
large overabundance of s-process elements.
Since the other objects in this small group of carbon-rich PPNs
all show the 21~$\mu$m emission feature
\citep[except perhaps for IRAS~07430+1115;][]{hrivnak99},
it is likely that IRAS~06530$-$0213 does also.  The infrared spectrum of the
object existing in the {\it IRAS} Low-Resolution Spectrometer database
is too noisy to judge the presence of the 21~$\mu$m feature, and the
position of the object in the sky made it inaccessible to observation
with {\it ISO}.  Thus the presence of the expected 21~$\mu$m feature
and also the 30~$\mu$m feature seen in carbon-rich evolved stars
remain to be confirmed.

\acknowledgements
We thank Eric Bakker for obtaining the initial spectra of IRAS~06530$-$0213 
and David Yong for obtaining one set of spectra of HD~56126.
We are grateful to the staff of UKIRT for the support of their service
observing program.  J. Ouyang assisted in the reduction of the SQIID
data.  B.~J.~H. acknowledges support from the National Science
Foundation  (AST-9315107, AST-9900846) and NASA (NAG5-3346), and
B.~E.~R. acknowledges support from the Robert A. Welch Foundation.
This research has made use of the SIMBAD database, operated
at CDS, Strasbourg, France, and the NASA ADS service.

\clearpage

\begin{table}
\caption{Visible and Near-Infrared Photometry of IRAS~06530$-$0213}
\label{ptm}
\begin{tabular}{ccccc}
\hline
\hline
B   &  V  &R  & I  & Observation Date \\
\hline
16.40$\pm$0.03 & 14.06$\pm$0.04 & 12.69$\pm$0.04 & 11.44$\pm$0.02 & 1995 Sep
11\\
\hline
J  & H & K & L & Observation Date \\
\hline
9.46$\pm$0.02 & 8.87$\pm$0.02 & 8.43$\pm$0.02 & 8.0$\pm$0.3 & 1993 Nov 9, 10\\
\hline
\end{tabular}
\end{table}

\clearpage

\begin{table}
\caption{Atmospheric Parameters Determined from Spectroscopic Analysis}
\label{model-atm}
\begin{tabular}{cccccc}
\hline
\hline
Name & $T_{\rm eff}$ & log~$g$ & $\xi_{t}$ & [M/H] &  $V_{\rm r}$  \\
          &   (K)   & (cm~s$^{-2}$) & (km~s$^{-1}$) & & (km~s$^{-1}$) \\
\hline
IRAS~06530$-$0213 & 6900 $\pm$ 250 & 1.0 $\pm$ 0.5 & 4.5 $\pm$ 0.5 & -1.0 &
51.0 $\pm$ 1.0 (1997 Oct 17) \\
 & & & & & 50.4 $\pm$ 1.0 (2001 Dec 10) \\
HD~56126 & 7250 $\pm$ 250 & 0.50 $\pm$ 0.5 & 5.0 $\pm$ 0.5 & -1.0 & 83.3 
$\pm$ 0.5 (1996 Dec 23) \\
 & & & & & 86.2 $\pm$ 0.5 (2002 Nov 14) \\
\hline
\end{tabular}
\end{table}

\clearpage
\begin{table}
\caption{Summary of Elemental Abundances}
\label{abund}
\begin{tabular}{lrrrrrrrrrrr}
\hline
\hline
        &\multicolumn{5}{c}{\underline
{~~~~~~~~~~IRAS~06530$-$0213~~~~~~~~~~~}}&&\multicolumn{5}{c}
{\underline{~~~~~~~~~~~~~~~~HD~56126~~~~~~~~~~~~~~~~}}\\
Species &$n$ & log $\epsilon$(X)   & $\sigma$ & [X/H] & [X/Fe] &
&$n$& log $\epsilon$(X)   &$\sigma$& [X/H] & [X/Fe]\\
\hline
C~I   & 19  & 8.72\tablenotemark{a} & 0.17 & 0.33\tablenotemark{a} & 1.26\tablenotemark{a} & & 33 & 8.09\tablenotemark{a} & 0.15 & -0.30\tablenotemark{a} & 0.70\tablenotemark{a}  \\
N~I   & 6   & 7.97\tablenotemark{b} & 0.28 &  0.05\tablenotemark{a} & 0.98\tablenotemark{b} & & 7  & 7.87\tablenotemark{b} & 0.12 & -0.10\tablenotemark{b} & 0.90\tablenotemark{b}  \\
O~I   & 3   & 8.30 & 0.10 &  -0.39 & 0.54& & 3  & 8.40 & 0.10 & -0.29 & 0.71 \\
%Na~I  & ... & ...  & ...  & ...   & ... & & 2  & 5.65 & ...  & -0.68 & 0.54 \\
Mg~I  & 2   & 6.73 & 0.17 & -0.85 & 0.08& & 5  & 6.76 & 0.13 & -0.82 & 0.18 \\
Si~I  & 5   & 6.90 & 0.11  & -0.65 & 0.27  & &3  & 6.78 & 0.11 & -0.77 & 0.23 \\
S~I   & ... & ...  &  ... & ...   & ... & & 2  & 6.49 & 0.10 & -0.72 & 0.28 \\
Ca~I  & 4   & 5.63 & 0.11 & -0.73 & 0.20& & 4  & 5.35 & 0.17 & -1.01 &-0.01 \\
Ca~II & 4   & 5.62 & 0.13  & -0.74 & 0.18& & 3  & 5.43 & 0.25   & -0.93 & 0.07 \\
Sc~II & ... & ...  & ...  & ...   & ... & & 4  & 2.41 & 0.07 & -0.76 & 0.24 \\
Ti~II & 2 & 4.15  & 0.08  & -0.87   & 0.05 & & 10 & 3.97 & 0.12 & -1.05 &-0.05 \\
Cr~II & 3 & 4.87  & 0.11  & -0.78   & 0.14 & & 14 & 4.76 & 0.09 & -0.91 & 0.09 \\
Fe~I  & 11  & 6.60 & 0.11 & -0.90 & 0.02& & 40 & 6.53 & 0.12 & -0.97 & 0.03 \\
Fe~II & 7   & 6.57 & 0.18 & -0.93 &-0.02& & 15 & 6.46 & 0.17 & -1.04 & -0.04 \\
Y~II  & 5   & 3.02 & 0.20 & 0.78  & 1.70& & 10  & 2.85 & 0.15 &  0.61 & 1.61 \\
Zr~II & 1 & 3.09  & ...  & 0.49   & 1.42 & & 8  & 3.02 & 0.15 &  0.42 & 1.42 \\
Ba\,{\sc ii}& 3  & 4.78 & 0.07 & 2.65 & 3.58\tablenotemark{c}& & 1& 2.45 & ... & 0.32 & 1.32 \\
La~II & 4   & 2.61 & 0.12  & 1.39  & 2.32& & 7  & 2.09 & 0.15 &  0.87 & 1.87 \\
Ce~II & 10   & 2.78 & 0.19 & 1.23  & 2.16& & 17 & 2.02 & 0.16 &  0.47 & 1.47 \\
Nd~II & 10   & 2.53 & 0.16 & 1.03  & 1.95& & 15 & 2.14 & 0.15 &  0.64 & 1.64 \\
Sm~II & ... & ...  &...   & ...   & ... & &  5 & 1.26 & 0.11 & 0.25 & 1.25 \\
Eu~II & ...   & ... & ...  & ...  & ... & & 1  & 0.03 & ...  & -0.48 & 0.52 \\
\hline
\end{tabular}
\tablenotetext{a}{Does not includes non-LTE correction of -0.3 dex; see text for details.}
\tablenotetext{b}{Does not includes non-LTE correction of -0.4 dex; see text for details.}
\tablenotetext{c}{Abundance is derived using very strong
 (W$_{\lambda}$ $>$ 400\,m\AA) Ba\,{\sc ii} lines.}
\end{table}
\clearpage

\begin{table}
\caption{Summary of Photospheric and Circumstellar Velocity Components}
\label{vel}
\begin{center}\scriptsize
\begin{tabular}{lccccccccccc}
\hline
\hline
 &\multicolumn{5}{c}{\underline{~~~~~~~~~~~~~~~~~~~IRAS~06530$-$0213~~~~~~~~~~~~~~~~~~~}}&
  &\multicolumn{5}{c}{\underline{~~~~~~~~~~~~~~~~~~~~~~~~HD~56126~~~~~~~~~~~~~~~~~~~~~~~~}} \\
 &\multicolumn{2}{c}{\underline{~~~~~~1997 Oct 17~~~~~~}}&
  &\multicolumn{2}{c}{\underline{~~~~~~2001 Dec 10~~~~~~}}&
  &\multicolumn{2}{c}{\underline{~~~~~~1996 Dec 23~~~~~~}}&
  &\multicolumn{2}{c}{\underline{~~~~~~2002 Nov 14~~~~~~}} \\
 & V$_{r}$ & V$_{\rm exp}$ & & V$_{r}$ & V$_{\rm exp}$ &
   & V$_{r}$ & V$_{\rm exp}$ & & V$_{r}$ & V$_{\rm exp}$ \\  
 Component & (km s$^{-1}$) & (km s$^{-1}$) & 
  & (km s$^{-1}$) & (km s$^{-1}$) &
  & (km s$^{-1}$) & (km s$^{-1}$) & & (km s$^{-1}$) & (km s$^{-1}$) \\
\hline
Phot (atomic lines)& 51.0$\pm$1.0  & ... & & 50.4$\pm$1.0  & ... & 
 & 83.3$\pm$0.5 & ... & & 86.2$\pm$0.5 & ... \\
System (CO)        & 50\tablenotemark{a} & ...  & & 50\tablenotemark{a} & ...  &
 & 87\tablenotemark{b} & ... & & 87\tablenotemark{b} & ...  \\
CSE (CO)             & ...  & $\sim$10\tablenotemark{a} & 
  & ...  & $\sim$10\tablenotemark{a} &
  & ...   & 10.2\tablenotemark{b} & & ...   & 10.2\tablenotemark{b} \\
CSE (C$_{2}$)      & 36.0$\pm$0.6 & 14  &  & 36.4$\pm$0.8 & 14  & 
  & 77.3$\pm$0.5 & 10 & & 76.6$\pm$1.0 & 11\\
CSE (CN)             & 37.4$\pm$1.5 & 13  & &  37.4$\pm$1.0 & 13  &
  & 77.2$\pm$0.5 & 10 & & 76.1$\pm$1.0 & 11\\
CSE (Na~I)            & ... & ... & & ... & ... & 
  & 74.0$\pm$0.5 & 13 & & 76.5$\pm$1.0 & 11\\
CSE (K~I)             & 34.0$\pm$1.0 & 16  & & ... & ... &
  & 75.5$\pm$1.0 & 12 & & 76.1$\pm$1.0 & 12 \\
\hline
\end{tabular}
\end{center}
\tablenotetext{a}{\citet{hu94}, transformed from $V_{\rm lrs}$=33 km~s$^{-1}$.}
\tablenotetext{b}{\citet{omont93}, transformed from $V_{\rm lrs}$=73 km~s$^{-1}$.}
\end{table}

\clearpage

\end{document}